\documentclass[aps,pra, twocolumn, showpacs]{revtex4-2}
\usepackage{amsmath}
\usepackage{amsfonts}
\usepackage{amssymb}
\usepackage{fancyhdr}
\usepackage{color}
\usepackage[utf8]{inputenc}
\usepackage{mathtools}
\usepackage{graphicx}
\usepackage{float}
\usepackage{physics}
\usepackage{bbold}
\usepackage{url} 
\usepackage{color}
\usepackage{cancel}
\usepackage{array} 
\usepackage{latexsym}
\usepackage{mathrsfs}
\usepackage{enumitem}
\DeclareGraphicsExtensions{.pdf,.png,.jpg}
\usepackage{array}
\usepackage{verbatim}
\usepackage{booktabs}
\usepackage{yfonts}
\usepackage{graphicx}
\usepackage[format=plain,justification=centerlast]{caption}
\usepackage[format=plain,justification=centerlast]{subcaption}

\usepackage[title]{appendix}
\usepackage[colorlinks=true, urlcolor=blue,citecolor=blue,linkcolor=blue]{hyperref}

\begin{document}

\title{Superresolving optical ruler based on spatial mode demultiplexing for systems evolving under Brownian motion} 

\author{Konrad Schlichtholz}
\email[]{konrad.schlichtholz@phdstud.ug.edu.pl}
\affiliation{International Centre for Theory of Quantum Technologies (ICTQT),
University of Gdansk, 80-308 Gdansk, Poland}

\begin{abstract}
The development of superresolution techniques, i.e., allowing for efficient resolution below the Rayleigh limit, became one of the important branches in contemporary optics and metrology. Recent findings show that perfect spatial mode demultiplexing (SPADE) into Hermite-Gauss modes followed by photon counting enables one to reach the quantum limit of precision in the task of estimation of separation between two weak stationary sources in the sub-Rayleigh regime. In order to check the limitations of the method, various imperfections such as misalignment or crosstalk between the modes were considered. 

Possible applications of the method in microscopy call for the adaptive measurement scheme, as the position of the measured system can evolve in time, causing non-negligible misalignment. In this paper, we examine the impact of Brownian motion of the center of the system of two weak incoherent sources of arbitrary relative brightness on adaptive SPADE measurement precision limits. The analysis is carried out using Fisher information, from which the limit of precision can be obtained by Cram\'er-Rao bound. As a result, we find that Rayleigh's curse is present in such a scenario; however, SPADE measurement can outperform perfect direct imaging. What is more, a suitable adjustment of the measurement time between alignments allows measurement with near-optimal precision. 

\end{abstract}

\maketitle
\section{Introduction}
Estimation of the distance between two light sources is an important subject in microscopy and astronomy. For separations that satisfy Rayleigh's criterion \cite{Rayleigh}, which requires them to be at least as large as the width of the point spread function resulting from diffraction, one of the prominent methods for resolving the distance is direct imaging. However, while smaller separations can be estimated with direct imaging \cite{goodman1985}, this task becomes harder with decreasing separation \cite{BETTENS199937,Ram4457}. This phenomenon is known as Rayleigh's curse. 

Various superresolution techniques have been proposed to overcome the limitations of direct imaging \cite{Hell:94,Klar8206,Hsu_2004,Delaubert:06,Helstrom,Tsang2019}, and using the tools of quantum metrology \cite{HelstromQD,Holevo,ParisQE,Giovannetti} one can identify methods that reach the maximal precision allowed by quantum mechanics. In particular, a method that is quantum optimal was recently proposed, that is, photon counting after perfect spatial mode demultiplexing (SPADE) into Hermite-Gauss modes \cite{Tsang1,Tsang2}. Furthermore, proof-of-principles experiments utilizing SPADE have already been conducted \cite{Paur:16,Yang:16,Tang:16,Tham}. and the method is a field of active experimental research \cite{Boucher:20,Santamaria:23,Santamaria:24,Rouviere:24}. 

The astrophysical applications of SPADE \cite{superresolution_starlight_Tsang_2019,hypothesis_testing_exoplanets_2021,Hyp_1, Schlichtholz:24} like, e.g., exoplanet detection, are one of the most studied. However, another branch of applications, microscopy, still needs vast development. When considering applications, one necessarily has to think about imperfections in the implementation of the method \cite{Tsang1,Len,Manuel,Giacomo,Linowski}. For applications in microscopy where, in many cases, the sources are in motion, causing a non-negligible misalignment, the necessity of an adaptive scheme that would align the apparatus during the measurement was pointed out \cite{Tsang1}. However, the system evolves between alignments when the photons are counted. This evolution, which in most cases is a passive \cite{Brown,Einstain,Libchaber}  or active \cite{Bechinger,FODOR2018106,Basu} Brownian motion, varies the probabilities of detecting photons in specific modes. Thus, it should be taken into account in the measurement scheme to ensure the correct estimation of the separation.

In this paper, we show how to account for passive Brownian motion in an adaptive measurement scheme based on SPADE for the estimation of the distance between two weak incoherent point light sources where the center of the measured system evolves under Brownian motion. This scenario corresponds to using this method as an ``optical ruler'' that allows for, e.g., intramolecular distance estimation (which is one of the important applications in microscopy \cite{Ruler}) where the source could be intramolecular or realized as attached fluorophores. Furthermore, we show that it is crucial to consider Brownian motion in the analysis, as it results in the reappearance of Rayleigh's curse, which, however, can be circumvented by proper scaling of the time between realignments. In addition, we show that even in the regime where Rayleigh's curse is present, SPADE can outperform perfect direct imaging.

\section{System and measurement scenario}
The system under consideration is composed of two incoherent point light sources with a fixed distance $d$ between them and arbitrary relative brightness. The sources in the system are weak in the sense that measured photons follow the Poisson distribution. The system is placed in some solvent characterized by a diffusion coefficient $D$. The solvent forces a Brownian motion on the system, resulting in a displacement of the center of the system from the starting point and a change in the orientation of an axis passing through the sources after a finite time $t$.

Let us introduce a few assumptions about the measurement scheme. First, one can filter out photons that originated from the system from photons that come from the solvent. Measurement is carried out in the imaging plane, say $x, y$, and thus only projection of the system into this plane is relevant for distance estimation. 
The imaging plane of the measurement apparatus is at a distance sufficiently large from the center of the system that the impact of the Brownian motion in the $z$ axis is negligible.
Photon detectors for all modes are perfect. 

The measurement procedure starts with a quick estimation of the center of the system and aligning the measurement apparatus with the center. In the next step, the apparatus performs spatial mode demultiplexing of the image in the Hermite-Gauss basis $\lbrace u_{nm}(\vec{r}) \rbrace$ centered, which is assumed to be at the origin of the system at time $t=0$ and counts photons in particular modes for some finite time $T$. After that, the procedure is repeated. 
\begin{figure}[!t]
    \centering
    \includegraphics[width=0.49\textwidth]{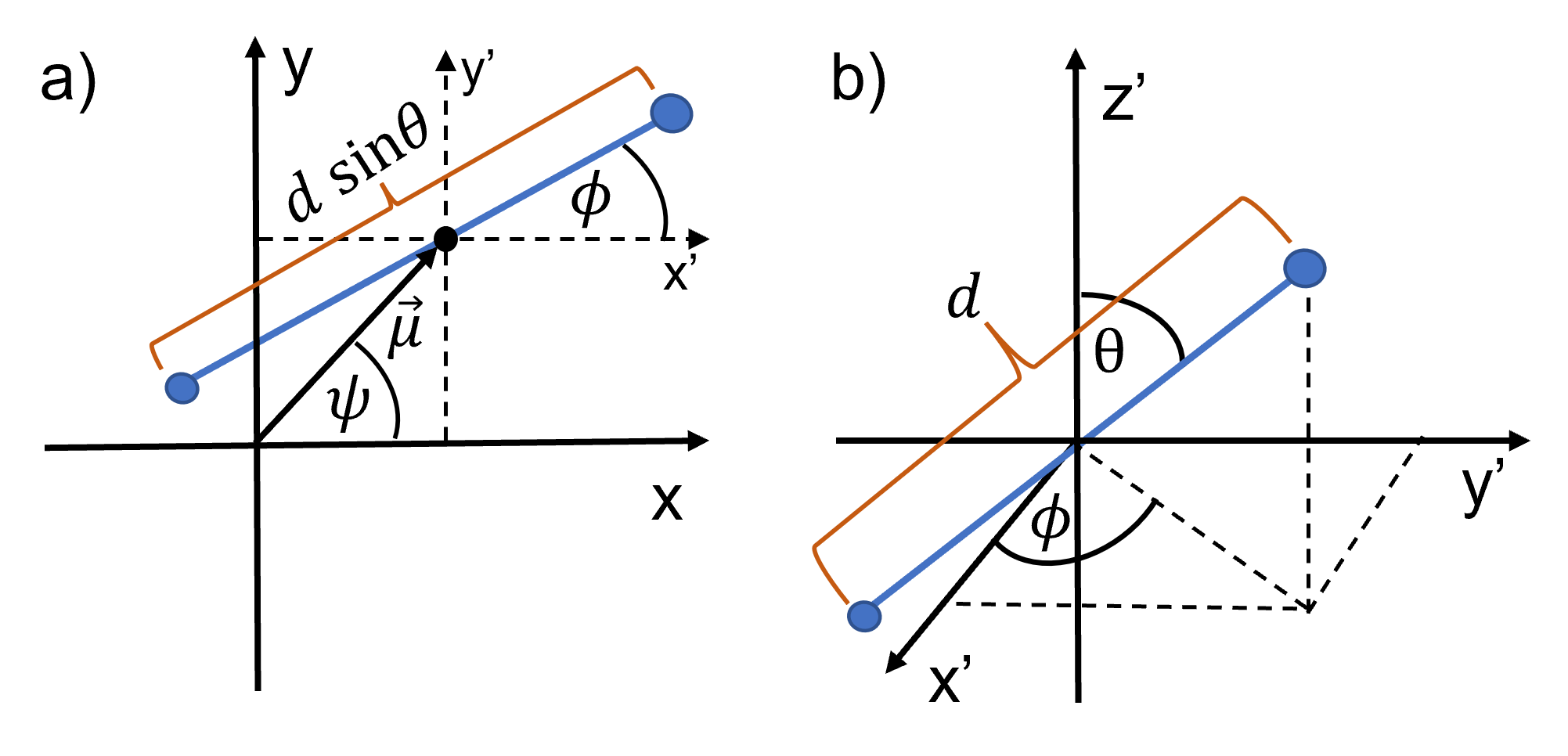}
    \caption{a) Projection of the system of two sources into $x,y$ plane of the imaging system. The center of the system is misaligned by $\vec{\mu}$. b) The system of two sources in a misaligned coordinate system $x',y',z'$. Two angles $\phi\in [0,2\pi)$ and $\theta\in [0,\pi]$ characterize the orientation of the system. }
    \label{fig:system}
\end{figure}
Due to the Brownian motion, each photon can be emitted from the system characterized by different misalignment vector $\vec{\mu}=(\mu\cos{\psi},\mu\sin{\psi})$ and orientation angles $(\phi,\theta)$ (see FIG. \ref{fig:system}). Note that for different $\theta$ the distance between sources in the projection into the imaging plane varies. Thus, the location of the sources is described by two vectors $\vec{r_i}=\vec{\mu}+(-1)^i(d\sin{(\theta)}/2)(\cos{\phi},\sin{\phi})$. A good approximation of the spatial distribution of the field after going through a diffraction-limited imaging apparatus is given by two overlapping Gaussian profiles centered on the locations of the sources $u_{00}(\vec{r}-\vec{r}_i)$ where $u_{00}(\vec{r})=\sqrt{2/(\pi w^2)}\exp{-r^2/w^2}$  \cite{goodman1985} coincides with $(0,0)$ Hermite-Gauss basis function. Let us denote by $\hat\rho(\vec{\mu},d,\phi,\theta)$ the density matrix of the field that comes from the misaligned system. 

We recall that for incoherent superposition of two weak thermal or coherent light sources at fixed positions $\vec{r_i}$ measurement can be interpreted as a sequential measurement on multiple copies of the single-photon state $\hat\rho(\vec{r_i})$ \cite{Linowski}. Furthermore, the probability of detecting the photon in the $(n,m)$ Hermite-Gauss mode is determined by the overlap integrals:
\begin{equation}
 f_{nm}(\vec{r}_i)=\int_{\mathbf{R}^2} d\vec{r}u_{nm}^*(r)u_{00}(\vec{r}-\vec{r}_i)
\end{equation}
and is given for our state $\hat\rho(\vec{\mu},d,\phi,\theta,\nu)$ by:
\begin{equation}
    p(nm|\vec{\mu},d,\phi,\theta,\nu)=\nu| f_{nm}(\vec{r}_1)|^2+(1-\nu)| f_{nm}(\vec{r}_2)|^2,
\end{equation}
where $\nu \in (0,1)$ stands for the relative brightness of the sources.
Furthermore, in \cite{rotations} method of considering separation estimation between sources undergoing motion on fixed trajectories was presented. 
We adapt this method to calculate probabilities of measuring the photon in $(n,m)$ mode in the case of Brownian motion (see Appendix \ref{app:stoch} for additional details). As the measurement scheme is repeated many times and motion is a random walk, our system is not described by some specific trajectory, but rather by some probability distribution of the misalignment vector arising from the two-dimensional Brownian distribution and the fully random distribution of orientation. Thus, at any given moment of time $t$ during our measurement repetition system is effectively the mixed state $\hat\rho_B(t,d,\nu)$ being a statistical mixture of states $\hat\rho(\vec{\mu},d,\phi,\theta,\nu)$ approximately given by:
\begin{multline}
   \hat\rho_B (t,d,\nu)=\\ \int_{\mathbf{R}^2} d\vec{\mu}\int_{\Omega} d\overline{\Omega} \hat\rho(\vec{\mu},d,\phi,\theta,\nu) \frac{1}{4 \pi D t}\exp{\frac{\mu^2}{4D t}},
\end{multline}
where we have used the formula for two-dimensional Brownian probability distribution of the center of the system \cite{Nelson}, and $d \overline{\Omega}=(d\phi d\theta\sin{\theta})/4\pi$ for which integration is over all orientation angles.
As the measurement cycle has a finite duration $T$ and there are no privileged points in time for measuring a photon, we have to make a time average of $\hat\rho_B (t,d)$ resulting in the state:
\begin{equation}
    \hat\rho_B (d,T,\nu)=\frac{1}{T}\int_0^Tdt\hat\rho_B (t,d,\nu).\label{rhoB}
\end{equation}
As a consequence, in our scenario, the probability of measuring a photon in mode $(n,m)$ is given by (for exact formulas, see Appendix \ref{app:B}):
\begin{multline}
      p(nm|d,T)=\\ \int_0^Tdt\int_{\mathbf{R}^2} d\vec{\mu}\int_{\Omega} d\overline{\Omega} p(nm|\vec{\mu},d,\phi,\theta,\nu)\frac{\exp{\frac{\mu^2}{4D t}}}{4\pi T D t}.\label{rhob}
\end{multline}
  We stress that we have found that these probabilities are independent from $\nu$. Let us introduce the following unit-less parameter $\tau=D T/w^2$. One can calculate variance $Var(\mu)=(2-4\pi/9) w^2 \tau$ using probability distribution from (\ref{rhob}) and thus $\sqrt{\tau}$ determines the spread of the system during the measurement.
\section{Precision limit of distance estimation}
Let us recall that the bound for the uncertainty of a distance estimation with an unbiased estimator is given by the Cram\'er-Rao bound \cite{Kay}:
$
\Delta d\geq1/\sqrt{N F(d)},
$
where $F(d)$ is the Fisher information (FI) per measured photon, which in the case of Poissonian photodetection is given by \cite{Chao:16}:
\begin{equation}\label{eq:FI_definition}
    F(d) = \sum_{n,m=0}^M \frac{1}{p(nm|d)} \left( \frac{\partial}{\partial d} p(nm|d) \right)^2,
\end{equation}
where $M$ denotes maximal index of Hermite-Gauss mode which can be distinguished by the measured apparatus considered.
One can obtain quantum FI $F_Q(d)$ through maximization of FI with respect to all physically possible measurements \cite{Kay}. It was shown \cite{Tsang1} 
that for distance estimation $F_Q(d)=w^{-2}$ and that it is achievable in the limit of $M\rightarrow\infty$ with measurement where the Hermite-Gauss modes are centered at the origin of the system. Moreover, for $d/2 w\ll 1$ and $M=1$ such measurement approaches the quantum limit with decreasing $d$.

The requirement that estimated $d$ must be greater than $\Delta d$ to have meaning results in the definition of the minimal resolvable distance \cite{Manuel} given by the solution of $d_{\min} = 1/\sqrt{N F(d_{\min})}.
$

\section{Asymptotic Fisher information in the case of Brownian motion}
Let us analyze Fisher information for our measurement scheme which we denote by $F_{HG}(d,\tau)$. We are mostly interested in the small distances regime  $x:=d/2w\ll 1$ thus, we focus our discussion on the dominant terms contributing to FI resulting from modes with $n,m\leq1$ (we calculate FI with $M=1$). 
Let us start by defining two timescales for our problem.  We refer to case $\sqrt{\tau}\ll x$ as a short timescale and to $x\ll \sqrt{\tau}$ as a long timescale. This is motivated by the fact that,  timescales should be defined in relation to separation as time of repetition $T$ determines the scale of a misalignment through $\sqrt{\tau}$. 

In order to analyze the asymptotic behaviour of FI, let us additionally assume that $\sqrt{\tau}\ll1$. This allows us to expand FI into a power series in $x$ and $\sqrt{\tau}$. Such an expansion can depend on the relation between the variables, as discussed in \cite{Linowski}. By expanding $F_{HG}(d,\tau)$ at first in $x$ and then in $\sqrt{\tau}$ for $x\ll\sqrt{\tau}$ and reversely for $x\gg\sqrt{\tau}$ we obtain:

\begin{align} \label{FiB}
    w^2 F_{HG}(d,\tau)\approx
    \begin{cases}
        \frac{2}{3}-\frac{2\tau}{x^2} & x\gg \sqrt{\tau}, \\   
       ( \frac{2}{9\tau}-\frac{43}{27})x^2 & x\ll \sqrt{\tau},
    \end{cases}
\end{align}
were for clearance we have taken only the dominant order in $x$ for two dominant orders in $\tau$. 
From this power series two main points can be stated. For short timescales, the impact of misalignment is minor, as the dominant term is constant and equal to $2/3$ which is the same as in the case of randomly oriented sources without misalignment reported in  \cite{rotations}. This claim is additionally supported by the fact that in the limit $\tau\rightarrow0$ full expression of $F_{HG}(d,\tau)$ results in the FI for the measurement with the same number of modes in the aforementioned scenario. For short timescales, the optimal scaling $d_{min}/w\sim N^{-1/2}$ is  approximately conserved. The second point is that for long timescales, optimal scaling is lost and replaced by the typical for direct imaging $d_{min}/w\sim N^{-1/4}$. This result shows that accounting for Brownian motion is necessary, as such motion can result in a significant drop in precision.  

\begin{figure}[!h]
    \centering
    \begin{subfigure}[b]{0.49\textwidth}
         \centering
         \includegraphics[width=\textwidth]{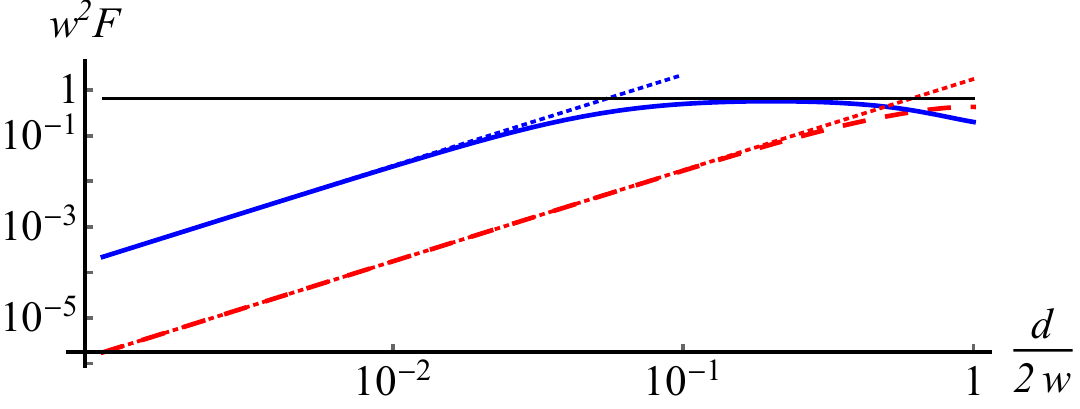}
     \end{subfigure} 
      \begin{subfigure}[b]{0.49\textwidth}
         \centering
         \vspace{-8pt}
         \includegraphics[width=\textwidth]{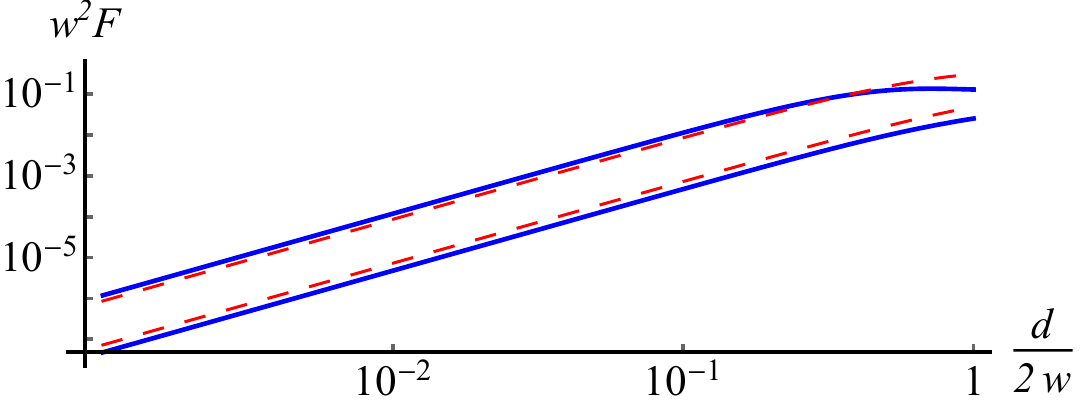}
     \end{subfigure}
    
        \caption{a) Comparison of Fisher information for SPADE measurement $F_{HG}(d,\tau)$ (blue curve) and for direct imaging $F_{DI}(d,\tau)$ (red dashed curve) and their asymptotic approximations (dotted curves of corresponding colors) for $\tau=0.001$. The black line corresponds to the optimal $2/3$. The advantage of SPADE in the presented case is up to two orders of magnitude. Also, transition between two timescales for SPADE is clearly visible. b) Comparison of Fisher information for SPADE and direct imaging (colors as in FIG. \ref{fig:FI_comp}a)). The top lines stand for $\tau=0.1$ and the bottom for $\tau=1$. In the first case, SPADE has a minor advantage even though $\sqrt{\tau}>0.29$. In the second case direct imaging has an advantage in this very longtime regime; however, both methods are comparable and perform poorly.  }
        \label{fig:FI_comp}  
\end{figure}

Despite reappearance of the Rayleigh curse, SPADE still can result in better resolution than direct imaging. In order to verify this, the power series of FI for perfect direct imaging measurement can be calculated (for details, see Appendix \ref{app:Direct}) resulting in the following expression for both timescales:
\begin{equation}
    w^2F_{DI}(d,\tau)\approx \frac{16 x^2}{9}-\frac{128 \tau  x^2}{9}+\frac{1792 \tau ^2 x^2}{27}.\label{FiD}
\end{equation}
 The coefficient of $x^2$ for direct imaging is smaller for $\sqrt{\tau}\lessapprox 0.29$ than for SPADE.  However, this does not necessarily mean that for higher values of $\sqrt{\tau}$ direct imaging performs better, since with increasing $\sqrt{\tau}$ more terms of the power series should be relevant and $F_{HG}(d,\tau)$ could be increased by measuring additional modes (see FIG. \ref{fig:FI_comp}b)). 
 

FIG. \ref{fig:FI_comp} shows the comparison of values of $F_{HG}(d,\tau)$ and numerical values of $F_{DI}(d,\tau)$ for small values of $x$ and $\tau\in \lbrace0.001,0.1,1\rbrace$. The results presented in this figure support the claims made above. On Fig: \ref{fig:FI_comp} a) continuous transition between timescales can be clearly seen. 
\section{Optimal scaling for $\tau$}
Important question is how to set the time of the measurement, to measure with the optimal scaling. Here, we consider the scaling of $\tau$ of the form $\sqrt{\tau}\rightarrow \kappa x^q$ with $\kappa\leq1$. Specifically, for $q=1/2,1,2$ FI has the following approximate form:
\begin{align} 
    w^2 F_{HG}(d,\sqrt{\tau}=x^q)\approx
    \begin{cases}
        \frac{2 x}{9\kappa^2}-\frac{2 x^2}{27 \kappa^4} & q=1/2, \\   
       \frac{2}{3(1+3\kappa^2)}-O(x^2) & q=1,\\
       \frac{2}{3}-\frac{8 x^2}{9} & q=2.
    \end{cases}
\end{align}
Note that for any $q>1$ we always end up in a short timescale regime for sufficiently small $x$. Thus, based on (\ref{FiB}) such scalings allow a measurement with near-optimal precision, however, with increasing $q$ convergence to the optimal scenario with decreasing $x$ becomes faster. For the case $q=2$ second term in the first line of (\ref{FiB}) becomes of the order $x^2$, and thus, for higher $q$, term constant in $\sqrt{\tau}$ and quadratic in $x$ will contribute more significantly resulting in an expansion of FI in second order in $x$ concurrent with the case $\tau\rightarrow 0$. From that follows that increasing $q$ over $2$ has a minor impact on the precision.

For $q<1$ we eventually get into the long timescale regime, but the higher $q$ the better the scaling can be achieved for minimal resolvable distance.

For the special case of $q=1$  and $\kappa=1 $ we are in the transition region between timescales. For such a scaling, an exact short timescale regime precision cannot be obtained. However, the constant term also appears, and as $\kappa$ decreases, it approaches the constant term for the short timescale. What is more, it reaches $90\%$ of optimal value already for $\kappa=1/3\sqrt{3}\approx 0.2$ and also converges to it optimally as $8x^2/9$ for small $\kappa$. 

These results show that the difficulty of the measurement increases with growing paste when decreasing $x$ as time in $\tau$ has to scale non-linearly with $x$ in order to measure with optimal scaling and setting lower $\tau$ requires more repetitions of measurement scheme to capture the same number of photons. Thus, setting $q=1$ seems optimal as it allows for the lowest-order scaling of time still achieving near optimal precision.

In a real-life scenario, one cannot scale $\sqrt{\tau}$ with some exact relation to $x$. However, the order of magnitude of $x$ can be estimated from some preliminary knowledge of the system or measurements.  Thus, our result allows one to estimate the optimal measurement time $T$. Also one can set the target $d_{min}$, and based on that  decide on $T$.

\section{Nonzero time of alignment}
In real-life scenario time of the alignment $t_a$ after the estimation of the center of the system has to be nonzero. During $t_a$, the system evolves, introducing some spread of the center of the system already at the beginning of the measurement. To account for this, we should perform time averaging  in (\ref{rhoB}) in bounds $[t_a,T]$ instead of $[0,T]$. Thus, in such a case probabilities of measuring photon in $(n,m)$ mode is given by:
\begin{equation}
    p(nm|d,T,t_a)=\frac{T\,p(nm|d,T)-t_a\, p(nm|d,t_a)}{T-t_a}.
\end{equation}
Assuming that time of alignment is short in comparison to time of measurement, i.e. $t_a\ll T$ we put $t_a=T/k$ where $k\gg1$. This modification results in approximate asymptotic FI:
\begin{align} \label{FiBt}
    w^2 F_{HG}(x,\tau,k)\approx
    \begin{cases}
        \frac{2}{3}-\frac{2(k+1)\tau}{k x^2} & x\gg \sqrt{\tau}, \\   
       ( \frac{2}{9\tau}+\frac{2}{9 k \tau}-\frac{43}{27}-\frac{23 }{27 k})x^2 & x\ll \sqrt{\tau},
    \end{cases}
\end{align}
were we have taken two dominant orders in $k$. The form of (\ref{FiBt}) is the same as (\ref{FiB}) with some small corrections. Thus, including $t_a$ results only in quantitative difference and the qualitative result is the same.
\section{Concluding remarks}
In summary, we have presented an adaptive scheme for the estimation of a distance between two weak incoherent sources that account for the Brownian motion of the system between the alignments of the measurement apparatus. Based on analytical considerations of Fisher information, we have shown that such a scheme employing spatial mode demultiplexing of the image into Hermite-Gauss basis, in theory, can perform better than its direct imaging counterpart in the limit of small distances even with measurement of a finite number of modes. This is both for the long timescale of the measurement where we have found that Rayleigh's curse is present and on the short time scale where SPADE performs with the optimal scaling of minimal resolvable distance. This result shows that Brownian motion between alignments cannot be simply neglected.  We show how the measurement time should be adjusted depending on the expected magnitude of the separation to perform the measurement, allowing for almost maximal precision. Finally, we have shown that a short nonzero time of alignment does not qualitatively impact our findings.

Development of such schemes of measurement is crucial in order to correctly implement SPADE measurements for molecular applications, where centers on the image plane of the measured systems can be in relatively quick motion in comparison with almost stationary stellar objects. As SPADE aspires to become a relevant high-precision method also for molecular applications, further refinements of schemes based on SPADE are necessary, for example, to avoid errors related to fitting data to a wrong underlying model of the method. Thus, also our scheme still looks for improvements, opening many possibilities for further research. First of all, the estimation of the center of the system and alignment of the measurement apparatus is burdened with some error. Thus the assumption of perfect alignment at the start of each run of the experiment should be discarded and replaced, for example, with some probability distribution for initial misalignment. What is more, the movement of some biological systems is governed by active Brownian motion rather than considered by us a passive one. Due to the highly non-Markovian behaviour of such systems on the short timescales \cite{Basu} they might require deeper consideration in order to properly apply SPADE based measurements for them. One could also consider the impact of crosstalk or dark counts. Extension to sources different from Poissonian could also be considered. However, not only could refinements of the underlying model be made, but maybe also improvements of the measuring scheme could be done, as, for example, introduction of some time binning of photon counts may result in some improvements.

\section{ACKNOWLEDGMENTS}
Project ApresSF is supported by the National Science Centre (No. 2019/32/Z/ST2/00017), Poland, under QuantERA, which has received funding from the European Union's Horizon 2020 research and innovation programme under Grant Agreement No. 731473. 
\bibliography{biblio}
\appendix
\section{Stochastic motion}\label{app:stoch}
In not all scenarios the exact trajectory is given, and dynamics is rather a stochastic process. In such a case, different trajectories have different probability of occurring. Thus, at each point of time, we have some probability density function $\textbf{p}(\vec{q_1},\vec{q_2},t)$ that the system of two sources is at specific coordinates $\vec q_1$ and $\vec q_2$. The measurement in our scenario could be generically seen as performed on the system in some mixed state:
\begin{equation}
    \hat \rho_s =\int d\vec{q_1}d\vec{q_2}\,\textbf{p}(\vec{q_1},\vec{q_2}) \hat\rho(\vec{q}_1,\vec{q}_2),
\end{equation}
where $\hat \rho(\vec{q}_1,\vec{q}_2)$ is the state of the sources in the static scenario for given coordinates $\vec q_i$ and $\textbf{p}(\vec{q_1},\vec{q_2})$ is some probability density function of coordinates.
Knowing the time-dependent probability distribution $\textbf{p}(\vec{q_1},\vec{q_2},t)$  one might try to calculate the $\hat\rho_s$ as a time average over measurement time $t_m$, effectively obtaining the following mixed state:
\begin{equation}
    \hat \rho_s \approx\frac{1}{t_m}\int_0^{t_m} dt\int d\vec{q_1}d\vec{q_2}\,\textbf{p}(\vec{q_1},\vec{q_2},t) \hat\rho(\vec{q}_1,\vec{q}_2). \label{stoch}
\end{equation}
However, only one specific path in our scenario is followed by the system during the time of measurement. Thus, it is not enough to collect a large number of photons to ensure the correctness of such an effective description.  It becomes necessary that $\textbf{p}(\vec{q_1},\vec{q_2},t)$ is periodic in time and there is no point in the allowed for the system part of the coordinate space from which some subspace of this space cannot be reached at any later time. What is more, in general, one also needs Markovianity of the evolution on some time scale much shorter than the time of measurement. This is, e.g., to prevent cases where the initial points determine some subclasses of trajectories with different probability distributions of positions as in such cases statistical properties of state evolving on such a subclass in an obvious way does not have to follow prediction given for the full ensemble of such subclasses when one starts from the probabilistic mixture of initial points. These requirements ensure us that any point in coordinate space is approached arbitrarily close by the system in the limit  $t_m\rightarrow \infty$ with expected frequency. Then equation (\ref{stoch}) becomes justified by the Ergodic theorem \cite{Ruelle1979} as then it is strict in the limit $t_m\rightarrow \infty$.  These requirements can be fulfilled, for example, by running the whole experiment $N_r$ times for some finite time of repetition (which determines a period) $T$ and starting always from the same initial state. In such a case, the limit of infinite time of measurement is replaced by the limit $N_r\rightarrow \infty$ as $t_m=N_r T$. Thus, with increasing number of experiment runs, the description given by (\ref{stoch}) becomes more accurate.

\section{Fisher information for direct imaging}
\label{app:Direct}

In this section, we present how Fisher information is calculated for the perfect direct imaging measurement. We recall that FI for the static scenario in the case of direct imaging is given by:
\begin{multline} \label{eq:FI_DI}
    F_{\textnormal{DI}}(d) =\\ \int_{\mathbf{R}^2} d\vec{r}\, \frac{1}{p(\vec{r}\,|\vec{\mu},d,\phi,\theta)} 
        \left( \frac{\partial}{\partial d} p(\vec{r}\,|\vec{\mu},d,\phi,\theta) \right)^2,
\end{multline}
where $p(\vec{r}\,|\vec{\mu},d,\phi,\theta)$ denotes probability density function of measuring the photon in location $\vec{r}$ where the location of the sources is given by some vectors $\vec{r_i}$ where $i=1,2$ and reads as follows \cite{Linowski}:
\begin{align} \label{eq:pdi}
    p(\vec{r}\,|\vec{\mu},d,\phi,\theta,\nu)=\nu |u_{1,00}|^2 + (1-\nu)|u_{2,00}|^2,
\end{align}
where $u_{i, 00}\coloneqq u_{00}(\vec{r}-\vec{r}_i)$ and $u_{00}(\vec{r})$ is the $(0,0)$ Hermite-Gauss basis function, in general, given by:
\begin{multline}
u_{nm}(\vec{r}=(r_x,r_y))= \frac{\exp{-(r_x^2+r_y^2)/w^2}}{\sqrt{(\pi/2)w^2 2^{n+m}n!m!}}\\ \times H_n\left(\sqrt{2}\frac{r_x}{w}\right)  H_m\left(\sqrt{2}\frac{r_y}{w}\right),  
\end{multline}
where $H_n()$ are Hermite polynomials and $w$ is the width of the point spread function.
In our scenario, this probability density is replaced by the following one:
\begin{multline}
        p(\vec{r}\,|d,\tau)=\\ \int_0^Tdt\int_{\mathbf{R}^2} d\vec{\mu}\int_{\Omega} d\overline{\Omega}  p(\vec{r}\,|\vec{\mu},d,\phi,\theta,\nu)\frac{\exp{\frac{\mu^2}{4D t}}}{4\pi T D t}.\label{eq:pDI}
\end{multline}
Note that $|u_{i,00}|^2$ are the same after integration for both $i$, and thus, the dependence on $\nu$ disappears. This is because replacing $\phi\rightarrow\phi +\pi$ we have $u_{1(2),00}\rightarrow u_{2(1),00}$ and the integration is over a uniform distribution of $\phi$. The resulting $F_{DI}(d,\tau)$ can be evaluated numerically. Probability (\ref{eq:pDI}) can be calculated approximately for $x\ll1$ and $\sqrt{\tau}\ll 1$. This can be done by noting that, due to the fact that orientation is fully random, our system is rotational invariant, i.e. $\forall{\vec{r}} \,\, p(\vec{r}\,|d,\tau)=p(|\vec{r}|\,|d,\tau) $. Thus, it is enough to calculate the probability for $\vec{r}=(r,0)$. Then, one can calculate the integral over $\mu$, which afterwards can be expanded into Taylor series in $x$ and $\sqrt{\tau}$. Finally, the rest of the integrals can be evaluated. In order to calculate $F_{DI}(d,\tau)$ one can expand the integrand in (\ref{eq:FI_DI}) into Taylor series in $x$ and $\sqrt{\tau}$ and after integration obtain:
\begin{equation}
    w^2F_{DI}(x,\tau)\approx \frac{16 x^2}{9}-\frac{128 \tau  x^2}{9}+\frac{1792 \tau ^2 x^2}{27}.
\end{equation}
Note that for $\tau\rightarrow0$ this expression goes to the $16x^2/9$ which is concurrent with the result for the system evolving under random rotations without Brownian motion reported in \cite{rotations}.

\section{Probabilities for SPADE measurement}
\label{app:B}
In this section we present probabilities from which the Fisher information for SPADE in Hermite-Gauss modes can be directly calculated.
 The probabilities under consideration are determined by overlap integrals:
\begin{equation}
 f_{nm}(\vec{r}_i)=\int_{\mathbf{R}^2} d\vec{r}u_{nm}^*(r)u_{00}(\vec{r}-\vec{r}_i)
\end{equation}
and are given by:
\begin{equation}
    p(nm|\vec{\mu},d,\phi,\theta)=\nu| f_{nm}(\vec{r}_1)|^2+(1-\nu)| f_{nm}(\vec{r}_2)|^2.
\end{equation}
The overlap integrals have the following form \cite{Giacomo}:

\begin{align}
\begin{split}
  &f_{nm}(\vec{r}_1)=e^{\frac{1}{2} \left(-x_{\theta}^2-\mu _w^2+2 x_{\theta} \mu _w \cos (\phi -\psi
   )\right)}\\&\frac{ \left(\mu _w \sin (\psi )-x_{\theta} \sin (\phi )\right){}^m
   \left(\mu _w \cos (\psi )-x_{\theta} \cos (\phi )\right){}^n}{\sqrt{m!
   n!}}, \label{f1}
\end{split}\\
  \begin{split}
 &f_{nm}(\vec{r}_2)=e^{\frac{1}{2} \left(-x_{\theta}^2-\mu _w^2-2 x_{\theta} \mu _w \cos (\phi -\psi
   )\right)}\\ &\frac{ \left(\mu _w \sin (\psi )+x_{\theta} \sin (\phi )\right){}^m
   \left(\mu _w \cos (\psi )+x_{\theta} \cos (\phi )\right){}^n}{\sqrt{m!
   n!}},  \label{f2}
   \end{split}
\end{align}
where $\mu_w:=\mu/w$ and $x_\theta:=x\sin{\theta}$. Using the following equation, one can compute the probabilities considered for the SPADE measurement:
\begin{multline}
      p(nm|d,\tau)=\\ \int_0^Tdt\int_{\mathbf{R}^2} d\vec{\mu}\int_{\Omega} d\overline{\Omega} p(nm|\vec{\mu},d,\phi,\theta)\frac{\exp{\frac{\mu^2}{4D t}}}{4\pi T D t}.
\end{multline}
The dependence on $\nu$ disappears analogously to the direct imaging case. We present probabilities for measuring a photon in mode $(n,m)$ for $n,m=0,1$:
\begin{widetext}
\begin{align}
    \begin{split}
         p(00|x,\tau)=&\frac{1}{12 \tau }2 \left(x^2 \left(\frac{\, _2F_2\left(1,1;2,\frac{5}{2};-\frac{x^2}{4 \tau +1}\right)}{4 \tau +1}-\, _2F_2\left(1,1;2,\frac{5}{2};-x^2\right)\right)+3 \log (4 \tau +1)\right),
    \end{split}\\
    \begin{split}
      p(10|x,\tau)=&\frac{1}{48\tau}\left(\frac{2}{x} \left(2 x^3 \left(\frac{\, _2F_2\left(1,1;2,\frac{5}{2};-\frac{x^2}{4 \tau +1}\right)}{4 \tau +1}-\, _2F_2\left(1,1;2,\frac{5}{2};-x^2\right)\right)+\frac{3 F\left(\frac{x}{\sqrt{4 \tau
   +1}}\right)}{\sqrt{4 \tau +1}}-3 F(x)\right)\right.\\
   &+6 \log (4 \tau +1)\Bigg),
    \end{split}\\
    \begin{split}
       p(11|x,\tau)=&\frac{1}{64\tau}\left(\frac{8 (4 \tau +1) x^2 \, _2F_2\left(1,1;2,\frac{5}{2};-x^2\right)-8 x^2 \, _2F_2\left(1,1;2,\frac{5}{2};-\frac{x^2}{4 \tau +1}\right)-12 (4 \tau +1) \log (4 \tau +1)}{12 \tau +3}\right.\\&\left.-\frac{\left((4 \tau
   +1) (32 \tau +7)+2 x^2\right) F\left(\frac{x}{\sqrt{4 \tau +1}}\right)}{(4 \tau +1)^{5/2} x}+\left(2 x+\frac{7}{x}\right) F(x)+\frac{1}{(4 \tau +1)^2}-1\right),
    \end{split}
\end{align}
\end{widetext}
where ${}_2F_2()$ stands for Hypergeometric PFQ function and $F()$ for the Dawson integral.

\end{document}